\documentclass[journal,comsoc]{IEEEtran}

\usepackage[T1]{fontenc}
\usepackage{amsmath}
\usepackage{xcolor}
\usepackage[cmintegrals]{newtxmath}
\usepackage{url}
\usepackage{hyperref}
\usepackage{graphicx}
\usepackage{cite}

\newcommand{\kETAL}    {{\em et~al.}}

\newcommand{\eg}	{{\em e.g.,~}}
\newcommand{\PEC}   {{PEC}}

\begin{document}
\title{Democratizing the Edge: A Pervasive Edge Computing Framework}%

\author{Reza Tourani, 
        Srikathyayani Srikanteswara, 
        Satyajayant Misra, 
      \\  Richard Chow, 
        Lily Yang, 
        Xiruo Liu, 
        and Yi Zhang 
\thanks{R. Tourani is with the Department of Computer Science, Saint Louis University. S. Srikanteswara, R. Chow, L. Yang, X. Liu, and Y. Zhang are with Intel Labs. S. Misra is with the Department of Computer Science, New Mexico State University.}}
\maketitle
\begin{abstract}
The needs of emerging applications, such as augmented and virtual reality, federated machine learning, and autonomous driving, have motivated edge computing--the push of computation capabilities to the edge. Various edge computing architectures have emerged, including multi-access edge computing and edge-cloud, all with the premise of reducing communication latency and augmenting privacy. However, these architectures rely on static and pre-deployed infrastructure, falling short in harnessing the abundant resources at the network's edge.

In this paper, we discuss the design of Pervasive Edge Computing (PEC)--a democratized edge computing framework, which enables end-user devices (\eg smartphones, IoT devices, and vehicles) to dynamically participate in a large-scale computing ecosystem. Our vision of the democratized edge involves the real-time composition of services using available edge resources like data, software, and compute-hardware from multiple stakeholders. We discuss how the novel Named-Data Networking architecture can facilitate service deployment, discovery, invocation, and migration. We also discuss the economic models critical to the adoption of PEC and the outstanding challenges for its full realization.
\end{abstract}

\section{Introduction}
\label{Introduction}
New breeds of applications such as autonomous driving, Industrial Internet of Things, and augmented and virtual reality (AR/VR) demand extensive computation, ultra-low latency communication, reliability, and improved security and privacy. To meet these needs, one solution is to delegate computation to more powerful surrogate machines at the network edge--called cyber-foraging~\cite{BalFliSat02}.
Numerous edge computing ecosystems have been proposed as solutions, including Cloudlets, fog computing, and Multi-Access Edge Computing (MEC)~\cite{TalSamMad17}. 
\begin{figure}[!t]
    \centering
    \includegraphics[width=0.8\columnwidth]{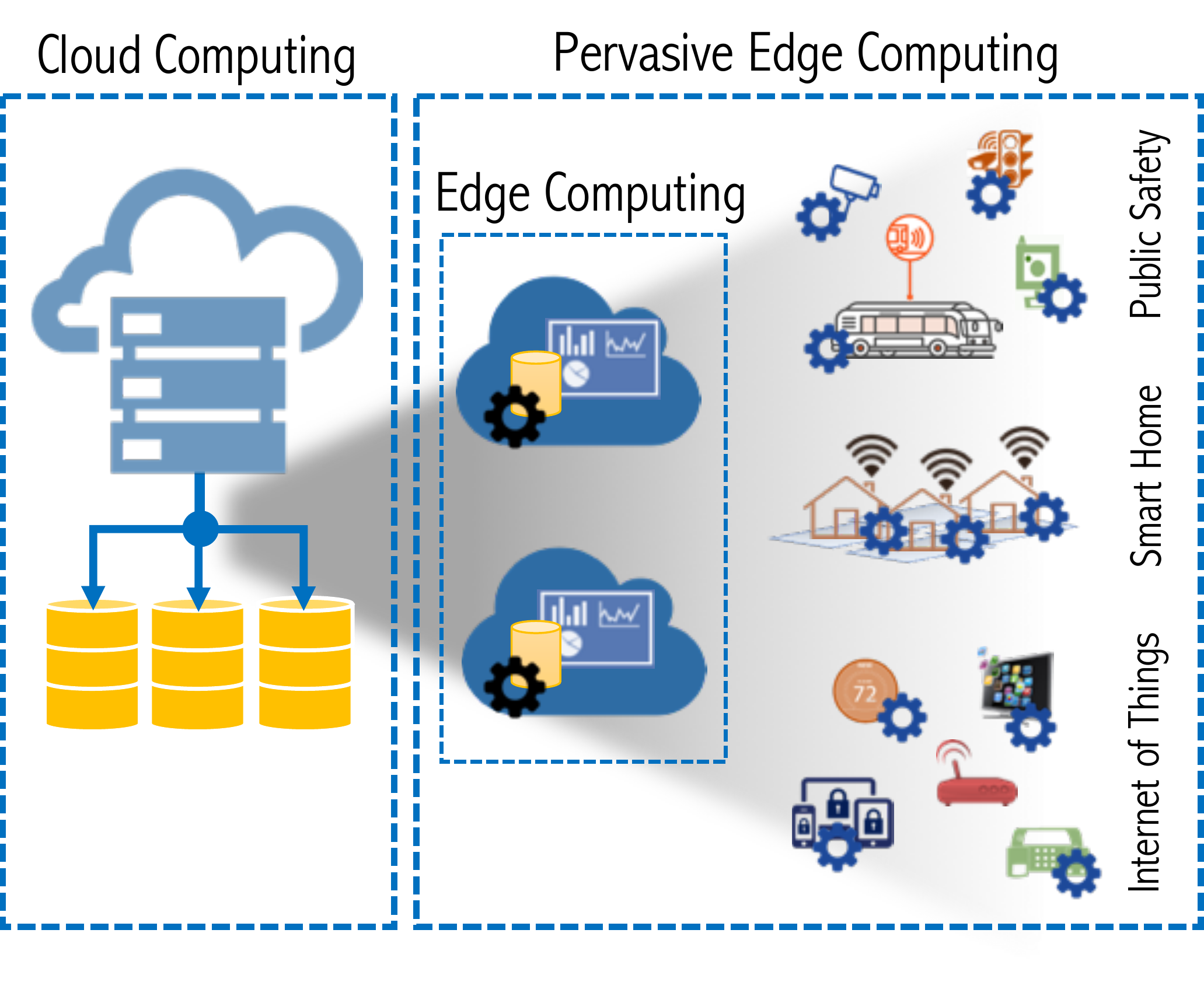}
    \caption{The computing ecosystem from cloud computing to pervasive edge computing, augmenting traditional edge computing with nearby resources.} 
    \label{fig:model}
    \vspace{-0.2in}
\end{figure}

Edge computing (EC) is considered a dynamic ecosystem due to high user mobility, unpredictability of wireless links, and dynamic resource availability. 
Thus, EC infrastructure must borrow and stitch together heterogeneous resources from multiple stakeholders--a missing piece in the existing edge ecosystems.
Consider an autonomous driving use-case, in which self-driving vehicles with multiple sensors need to create localized context from collected information. 
Given the several vehicles in proximity to each other, there is a potential for them to use each others' resources as well as infrastructure and end-users' devices for performing compute-intensive tasks. We refer the reader to~\cite{WanZhaLi19} for examples of applications. However, connections in this scenario would be short-lived, which makes effective service delivery and resource utilization challenging.
Our use-case goes far beyond the current use of vehicle-to-infrastructure communications based emergency applications, which only use low-bandwidth broadcast messages.   

Motivated by these challenges, we propose Pervasive Edge Computing (PEC)--a framework to augment existing EC architectures by provisioning end-user resources to enable markedly better performance (refer to Fig.~\ref{fig:model}). Harnessing end-user resources (hardware and software), considered in totality, can provide more and proximal resources to deploy services despite their low individual capacity. However, their erratic availability results in the PEC framework being significantly more dynamic than EC.  
Different from EC's orchestrated services and resources, PEC enables real-time composition of services leveraging dynamic end-user and infrastructure-based EC resources. 
%
However, scalable PEC framework requires efficient mechanisms for distributed service/resource discovery, service placement, and migration. 
To meet these requirements, we present the architectural design for the PEC framework, leveraging Named-Data Networking (NDN)~\cite{ZhaAfaBur14} as a communication architecture, which is a natural fit for distributed computing ecosystems~\cite{MtiTouMis18}. 


To illustrate NDN's advantages, consider the emerging applications in our use-case, such as using EC by the vehicles to search for license plates of cars searched by law enforcement~\cite{WanZhaLi19} or re-routing around an accident. 
%
Often, in these applications, the generated context (\eg an annotated accident scene) from the raw sensory data is useful for a group of users. Enabling context sharing in IP-based network, however, requires the application layer logic. In contrast, NDN can facilitate such context sharing through its descriptive content naming, semantic-based communication, and pervasive caching at the network layer~\cite{TouMtiMis19}. 
%

Our novel {\bf contributions} include:
{\it (i)} Design of a PEC framework that supports a dynamic multi-stakeholder ecosystem for seamless edge service delivery by utilizing all available edge resources, and general enough to incorporate future EC advancement.
{\it(ii)} Description of dynamic service orchestration, including service deployment, discovery, invocation, and migration in a multi-stakeholder ecosystem. 
{\it(iii)} Discussion of potential economic models of the multi-stakeholder edge.
{\it(iv)} Illustration of open PEC challenges and potential future research directions.
%
\section{Background and Preliminaries}
\label{Background}
\subsection{Edge Computing Frameworks} 
\label{MEC}
Numerous EC frameworks have been proposed. Liu~\kETAL~classified EC architectures into three types: Push from Cloud, Pull from IoT, and Hybrid Edge-Cloud Analytics; although, most frameworks in these categories rely on the cloud~\cite{liu2019survey}. Cloudlet and MEC fall under the first category, with MEC being one of the more standardized frameworks with clear APIs.
Paradrop, FocusStack, and SpanEdge are examples of the second category. SpanEdge reduces the latency in stream processing applications by placing the critical parts near the edge. Hybrid Edge-Cloud Analytics includes Amazon's Greengrass and Alibaba's LinkEdge. Greengrass allows users to place AWS lambda functions on IoT devices at the edge to reduce the latency for cloud access.

In this paper, we enrich the vision of EC by adding end-user resources that can contribute to an EC workload to the deployed infrastructure. This enables peer-based multiple end-users participation in the EC workload--{\em from the cloud to the crowd}. In PEC, we foresee real-time composition of services using software, data, and compute-hardware at the edge.

\subsection{Limitations of Existing Edge Computing Architectures} 
\label{Limitations}
The current EC architectures do not address some of the important challenges that will exist in a PEC. Here, we briefly discuss a few of them.

{\it Edge Resource Discovery:} Edge is a distributed and chaotic environment due to resource heterogeneity, traffic diversity/volatility, mobility, and varying link conditions. These characteristics result in a highly unpredictable ecosystem, which makes discovering resources/services a challenging task. 

{\it Service Orchestration:} 
The distributed nature of EC requires significant communication for obtaining the global view; high overhead if there is one controller, much worse if there are several controllers to guarantee fault tolerance. This results in depending on localized information availability and decision making based on partial information, potentially resulting in sub-optimal resource utilization decisions, which negatively impacts users' Quality of Experience.
%

%

{\it Seamless Service Handover:} Seamless service delivery for mobile users is a challenging task in existing EC architectures. Mechanisms need to be devised for migration of highly mobile users' services across multiple edge servers to assure uninterruptible service handover. This is particularly important as edge orchestrators allocate different services to certain servers for optimization purposes, which force some users to simultaneously connect to multiple distinct edge domains~\cite{TalSamMad17}.

{\it Security, Reliability, and Trust:} EC demands a fully distributed authentication mechanism to handle sections of infrastructure with limited connectivity to a centralized authentication server. Moreover, the multi-stakeholder nature of edge calls for a more nuanced and decentralized trust model.
%
\begin{figure}[!h]
    \centering
    \includegraphics[width=0.9\columnwidth]{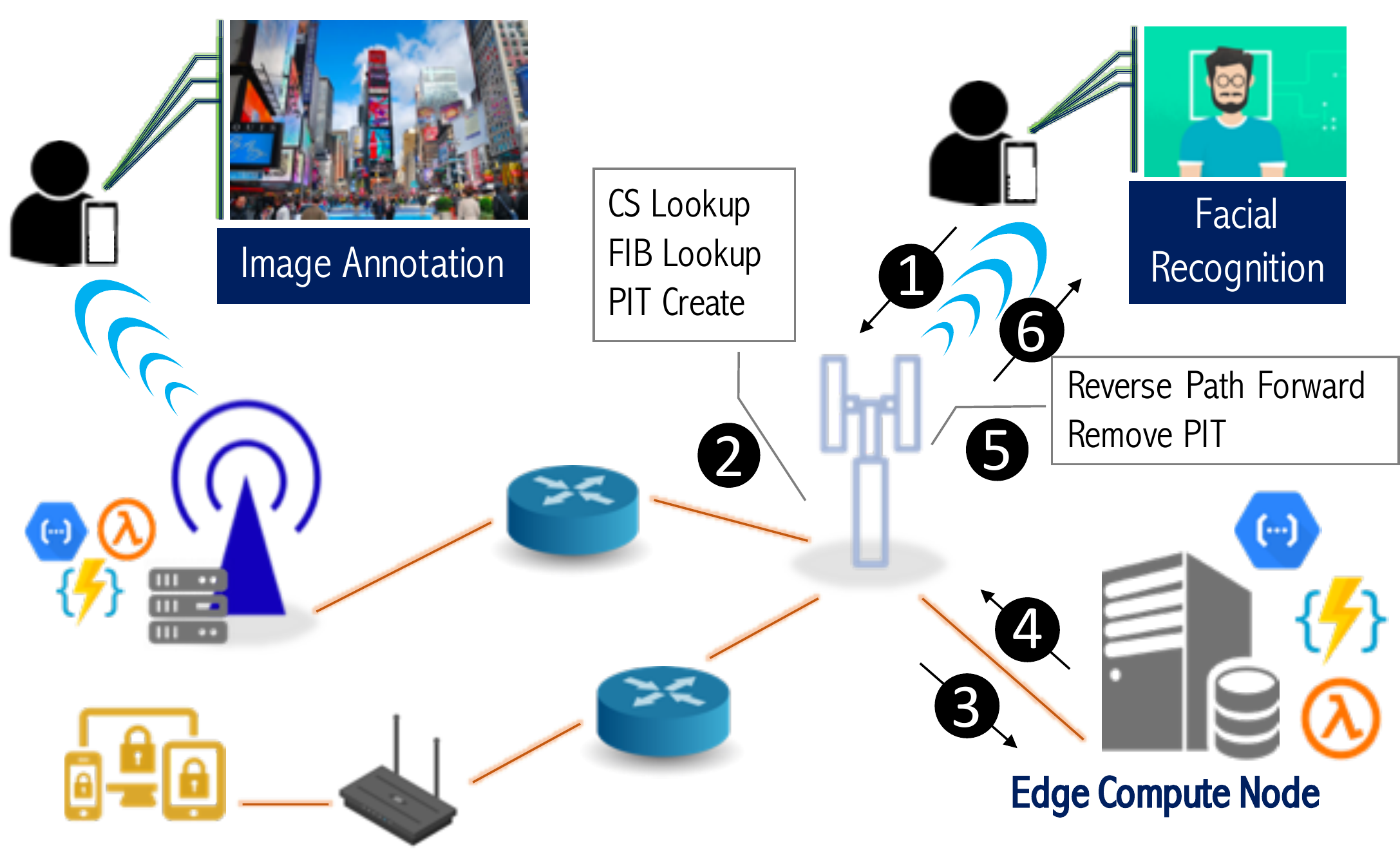}
    \caption{The Named-Data Networking communication flow, in which the response takes the request's reverse path.}
    \label{fig:compute}
\end{figure}

\subsection{Named-Data Networking}
\label{NDN}
%
The NDN architecture~\cite{ZhaAfaBur14} utilizes content naming, content caching, and name-based routing for content delivery. In NDN, entities are equipped with a content store (CS), a pending interest table (PIT), and a forwarding-information base (FIB). While the FIB corresponds to an IP forwarding table, the PIT keeps track of in-flight requests, allowing request aggregation and reverse path forwarding (Fig.~\ref{fig:compute}). Requesting data by name allows the intermediate entities to perform a CS lookup. On a failed CS lookup, the entity performs a PIT lookup for aggregation (Step 2 in Fig.~\ref{fig:compute}). 

Successful PIT lookup results in request suppression, in which the request will be dropped and the incoming interface will be stored. A failed PIT lookup results in creation of a PIT entry and request forwarding (towards the data source). The data takes the reverse path of the request to the user (Steps~4-6 in Fig.~\ref{fig:compute})). 
NDN forwarding strategy layer, different from routing, provides flexible packet forwarding (\eg multicast and broadcast).

%
Recent initiatives in NDN edge computing focus on placing functions in the network and executing them through virtual machines~\cite{SifKohSch14}, resource discovery~\cite{MtiTouMis18}, dynamic code execution on virtual machines~\cite{KroPsa17}, and Security-as-a-Service at the edge~\cite{TouBosMis19}. 
%
RICE~\cite{KroHabOra18}, an NDN-enabled EC framework, decoupled service invocation from result delivery to account for compute-intensive services. A similar approach has been employed in ICedge~\cite{MasMtiLee20}, which suggested mechanisms to leverage the existing computation (compute re-utilization).

%
\section{Multi-Stakeholders Edge Architecture}
\label{Architecture}
In this section, we introduce the \PEC's stakeholders (roles) and elaborate on their interactions in the context of our framework.

\subsection{Stakeholders}
%
Given PEC's fluid nature, an entity may have several concurrent roles. For instance, an end-user's device could be a user in need of an annotation service, while it could be a part of the infrastructure when it is responding to a crowd-sourced query. The typical stakeholders are:

\noindent {\bf User:} may be a smartphone, an IoT sensor, or an autonomous vehicle. Users generate raw data, which will be transmitted to a PEC server(s) for execution of edge applications to generate deeper context. 

\noindent {\bf Infrastructure:} is composed of all the hardware that can contribute to service executions, including Internet Service Provider's (ISP) hardware (\eg Comcast and AT\&T), private businesses' hardware for performing EC (\eg a Starbucks' wireless routers and traffic cameras), and end-user devices (\eg smartphones and vehicles). ISPs own the network from the cloud/core to the end-users and can rent out the hardware (\eg EC servers, smart WiFi access points, and Small cells) for PEC.

\noindent {\bf Orchestrator (Software Platform Provider):} is akin to cloud service providers (\eg Amazon or Google). These providers create an edge software orchestrator to be deployed on edge resources owned by the infrastructure provider to operate and manage the edge. At the back end, the edge orchestrator is connected to the cloud. Cloud Providers are already deploying EC software in private companies for data analytics (\eg Amazon Edge Lambda).
Aimed at democratizing the edge, our framework allows the end-user to be an orchestrator--an end-user can orchestrate services on its own devices and interact with a larger orchestrator (\eg Amazon) to bring in more resources. 

\noindent {\bf Edge Application (Service) Provider:} is an entity that leases edge resources, by interacting with the orchestrator to run the application on one/more VMs or containers. These applications are invoked by end-users as services. We consider two types of applications: {\it(i)} the unified applications such as AR/VR that are provided by known and trustworthy providers and {\it(ii)} unverified applications, which may be dynamically fetched and loaded into machines for execution upon users' requests. 
    
\begin{figure}[!t]
    \centering
    \includegraphics[width=0.9\columnwidth]{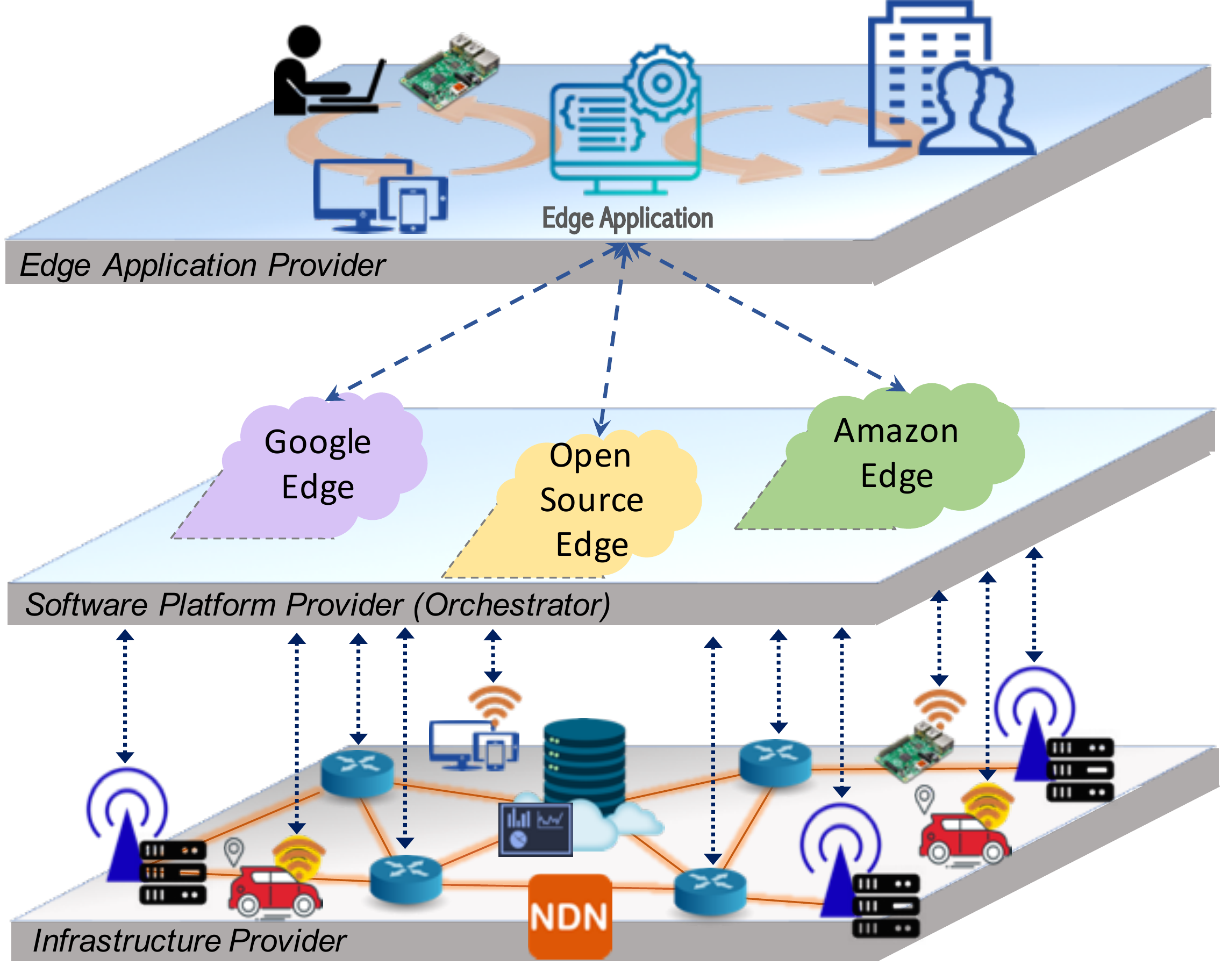}
    \caption{Our envisioned pervasive edge computing ecosystem along with various stakeholders distributed across the network.}
    \label{fig:architecture}
\end{figure}
%
\subsection{NDN-enabled Pervasive Edge Computing Framework}
\label{Design}
Our envisioned three-layer PEC framework (Fig.~\ref{fig:architecture}) uses NDN as the underlying networking substrate. The infrastructure layer (bottom layer) contains the edge/core networking infrastructure (includes routers and servers) with the geographically distributed computing resources. The orchestrator layer (middle layer) consists of both third-party open-source orchestrators and existing cloud service providers, with the orchestrator software running on the infrastructure. The third layer is composed of edge applications, provided by third party application developers or end-users. The interactions between the applications and users happen through the orchestrator, which hides the complexity and presents a well-defined substrate (through VMs or containers) for the applications.   
%

%
In PEC, the orchestrator leases hardware from the available resource pool to run VMs or containers on them. To onboard an end-user's device onto the computing pool, the device sends a join request via the access point it is connected to the PEC domain to the associated orchestrator (the orchestration domain that the access point operates in). Subsequently, the orchestrator instructs the associated access point to add a new interface towards the user's device for bidirectional flow. The onboarded device will then be considered an edge server and can participate in edge service execution.

We envision two approaches for edge service execution. In the conventional approach, the orchestrator provides VMs to the application providers for running edge applications. This is congruent with the existing EC architecture where the user's application runs on edge servers.
In the second approach, the orchestrator employs a serverless computing model to distribute computation in form of containers and data, allowing an application to be executed across multiple servers. Hence, enabling a coalition of constrained computing resources to perform similar computation as more powerful servers.

The task of selecting an edge server can be delegated to the network with the knowledge of available edge resources~\cite{MasMtiLee20}. Alternatively, the end-user may obtain the resource list to independently select the edge server. In the former, a biased router may unfairly redirect the requests toward preferred destinations (violating net-neutrality). In the latter, malicious end-users may mount a distributed denial of service (DDoS) attack by assigning consecutive compute-intensive tasks. These are potential areas to explore. In what follows, we discuss several aspects of our architecture.
%
%
\section{Service Orchestration: Deployment, Discovery, Invocation, and Migration}
\label{ServiceOrchestration}
\subsection{Service Deployment}
Upon a service request, orchestrators should deploy the service in the edge infrastructure. Services can be deployed in one PEC domains (one server) or across multiple PEC domains (multiple edge devices/servers operating under one or more orchestrators), depending on the service type, load status, and the geographical distribution of resources. Thus, we envision two approaches for service deployment.

\subsubsection{Orchestrated Service Deployment} 
A proportion of PEC's services would be pre-deployed, with the deployment informed by traffic, popularity, and demand models. This deployment aims to achieve a global optimal in efficient and effective resource utilization and users' experience. 
Orchestrated service deployment is managed by a control plane from the main data-center, which simplifies resources management. However, it introduces a single point of failure during network partitioning or cloud's unavailability. It also imposes significant communication overhead and will be slow in adopting edge dynamism. 
%

\subsubsection{Non-Orchestrated/On-demand Service Deployment}
PEC allows any device to become an application provider, resulting in independent pop-up services. In such scenarios, service deployment can be optimized in a local sense (in a local PEC domain). Given the fluctuations in the number of edge entities, it is almost impossible to schedule the services and the resources for global efficiency. Instead, localized controllers can be deployed that take into account the dynamics and marshal the edge resources for better scalability.
%
In this model, the autonomous PEC domains, with independent controllers (hierarchies of controllers), enable the orchestrators to fetch an absent service from each other. This model offers autonomy and resilience at the level of the orchestrator. However, it demands a greater effort in maintaining a large number of independent control planes.

\subsection{Service Discovery}
%
%
End-users entering a PEC domain should be able to seamlessly discover available services. In our design, the end-user sends a service discovery request into the network using a predefined name, which elicits a response containing the names of available services~\cite{MasMtiLee20}. If the requested service is available at the edge, the user utilizes the service name for the subsequent communication. 
Otherwise, the edge server decides whether to delegate the service execution to other edge servers or initiate the service migration from another edge server. Such a user-driven service discovery can be executed independent of service deployment approaches.
%

\subsection{Service Invocation}
For service invocation, the user sends a named request to the network, with the name including at least two components--the service name and the data name (for the edge server to fetch the data if missing). The edge server replies the service execution result if it does not incur high latency. The compute intensive service execution, however, requires the user to pull the computation result in a follow up communication.
We consider two service invocation scenarios: 
{\it 1)} Intra-Stakeholder Service Execution (non-cooperative stakeholders), and 
{\it 2)} Inter-Stakeholder Service Execution (cooperative stakeholders).

\subsubsection{Intra-Stakeholder Service Execution}
Considering non-cooperative stakeholders, on receiving a service execution request, the edge server has three options:
a) it has the capacity and executes the request; b) it does not have the requisite capacity and relays the request to another server within the same orchestrator domain; or c) none of the servers in the orchestrator domain have the capacity and the request has to be sent to the cloud.
%
For high latency service executions, the authors in~\cite{KroHabOra18} proposed using {\it thunk} names, which includes the processing server's identity, the requested service, and the service's state. Thus, on receiving a request, the edge server returns a thunk name and an estimated service execution latency, allowing the user to request the result after the expected waiting period. This approach can be used in our architecture as it is designed to allow new features to be plug-and-play.

\subsubsection{Inter-Stakeholder Service Execution}
Given financial incentives, PEC enables the cooperative stakeholders' orchestrators to provide services for each others' users. Similar to the telecommunication's roaming service, the edge server of a given orchestrator (without sufficient capabilities) delegates service execution to proximal edge server(s) in a neighboring orchestrator. For clarity, we say the designated edge server (selected by the user) delegates the service execution to a peering edge server (destination edge server) in another stakeholder domain.

In this scenario, the destination returns a thunk name to the designated edge server, which can either provide it to the user or generate another thunk name for the user and store the mapping between thunk names. The former allows the user to directly interact with the destination edge server while the latter creates a level of indirection. However, it allows the designated edge server to track the successful service execution by the peering stakeholder and hence provides robust accountability.

\subsection{Service Migration to Handle Mobility}
\label{migration}
%
%
Although NDN's connectionless nature facilitates seamless data delivery to mobile users, handling data upload and computation offloading from highly mobile users is still challenging~\cite{SchEmeMar19}. Thus, mechanisms needed for relocating the mobile user's ongoing service execution across edge servers--service migration--to minimize relocation time and interruptions; an area that has not been explored in NDN. 
%
%
For service migration, the destination edge server should decide the migration level depending on factors, such as the availability of guest operating system, requested service, and the estimated service execution time.
In the following, we briefly introduce three service migration levels.


\subsubsection{Result Migration}
is possible when the service execution in the source edge server is completed by the time the mobile user connects to the destination edge server. The authors in~\cite{SchEmeMar19} proposed a result migration mechanism, which results in low communication and computation overhead. 

\subsubsection{Instance Migration}
is the process of relocating a service's running instance by transferring the memory state of a VM (or a container) from its source to the destination edge server~\cite{WanXuZha18}. Instance migration can be performed either by duplicating all memory pages from the source to the destination while the VM instance is still running or suspending the VM instance and moving state data such as CPU state, register, and non-pageable memory to the destination~\cite{WanXuZha18}. Thus, it incurs moderate communication overhead, assuming the service is already running on the destination server.

\subsubsection{Service Migration}
transfers the complete service across edge servers when the destination does not possess the requested service. Service migration can be done as a complete VM (or container) migration. Depending on the destination server configuration, the VM might need to include the guest operating system (system kernel), required libraries, the executing service, and the user's instance. If the environment is provided, the migration only needs to relocate the service container and the user's running instance to the destination. 
%
%
\begin{figure}[!t]
    \centering
    \includegraphics[width=0.9\columnwidth]{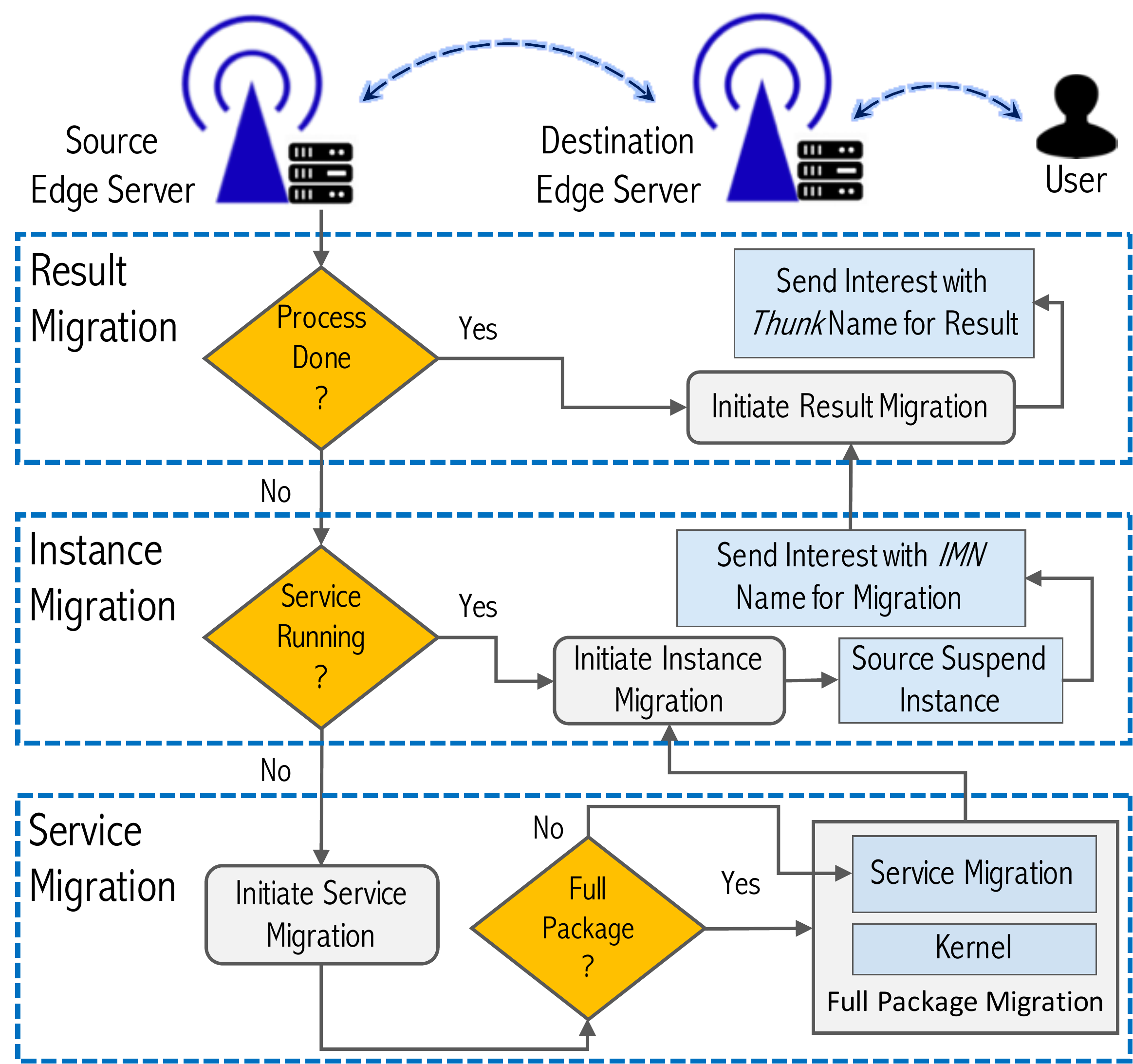}
    \caption{Illustration of interactions between the user, destination, and source edge servers resulting in the choice of the appropriate migration process by the source edge server.}
    \label{fig:migration}
\end{figure}

In Fig.~\ref{fig:migration}, we schematize the three levels of migration as a flowchart (self-explanatory). Leveraging NDN's naming can significantly simplify service migration process. For instance, the source edge server, upon receiving a service request, can assign a unique {\it instance migration name (IMN)} for the user's requested service--analogous to thunk name.
Upon the expiry of estimated service execution time, the user requests the result using the thunk and IMN. A static user who remains associated to the same edge server can use the thunk name to retrieve the service execution results. For a mobile user, the newly associated edge server (destination) can use the IMN sent by the user to initiate the migration process with its neighboring edge server (source).
%

%
%

\subsection{Re-use of Service}
%

%
In PEC, often users will request similar services. For instance, the users driving in the same direction of a highway may all request the annotation of an accident by providing their camera feed--although from slightly different angles as they all capture the same set of events.
NDN allows the edge servers and intermediate entities to intelligently detect similar offloaded tasks that share a given context, location, or objectives using the expressiveness of the names~\cite{MasMtiLee20}.
As a result, the edge server can reuse the result of the overlapped tasks from different users requesting the same service with similar data, resulting in significant savings.
%
Implementing such capability in an IP infrastructure will be exceedingly challenging, due 
to the lack of visibility of application information with the intermediate entities.
%
%
\section{Economics Models, Pricing, and Resource Accounting}
\label{Economic}
Edge computing presents economic model design challenges for both deploying the infrastructure and managing the pre- and post-processed data. In particular, in PEC, the majority of the high capacity infrastructure may be deployed by the infrastructure providers, namely cities, ISPs, or cellular network providers. Much of the orchestration software (software-as-a-service) may be deployed by software and cloud companies, such as Amazon and Microsoft. One of our primary design goals is to democratize the edge and keep barriers to entry low, enabling any entity/individual to enter and operate in the ecosystem to build and deploy edge applications for end-users. Further, a subscriber of, say, T-Mobile, can get service from an edge server running on the AT\&T network, or can even run his or her own PEC network that interfaces with other orchestrators to provide service.

Thus, a mechanism needs to be in place for AT\&T to provide the service while getting paid by T-Mobile (or the end-user PEC). There is need for a cost-sharing model between these entities and a means for money to flow between the entities involved in the chain. For these monetary exchanges, distributed ledger based technologies look promising. Distributed ledgers facilitate multi-party information exchange, asynchronous payment transmission, and auditing of the PEC economics. There have been some initial attempts at using blockchains for payment in the MEC~\cite{pan2018edgechain}.

Privacy can be another driver for new economic models. Edge servers collect and process end-user data, performing operations (\eg video annotation and overlaying of images). Recent policies, such as General Data Privacy Regulation (GDPR) and the California Consumer Privacy Act (CCPA), give users increased control over their data. An emerging model is for entities to remunerate the end-user for their data. 

These considerations suggest the possibility of an independent entity serving as a data/financial clearing house. Such an entity will be the conduit for all the communications, ensure traceability for the user data flows, and oversee the mechanism for user payments. Payments may happen with today's financial systems or on a crypto-currency network. Interesting trust and reliability questions arise as all this will only work if entities are trustworthy and follow protocols faithfully. Hence, mechanisms are needed to provide assurance of compliance to the privacy regulations or to prevent unauthorized sharing. We believe the technology solutions in this area need to develop together with innovations in the legal and regulatory space.          
%
%
\section{Challenges and Opportunities}
\label{Challenges}
%
Successful deployment of a PEC architecture introduces challenges and opens up opportunities for research. Here, we elaborate on a few these opportunities.

\subsection {Mobility} The PEC framework enables the edge server functionality to be deployed on moving platforms such as vehicles. Such deployments result in highly dynamic wireless link characteristics, which are difficult to model, estimate, and provision. Additionally, seamless service delivery to mobile devices requires fast service discovery, service invocation, and service migration among the local service providers. In cellular systems, mobility at the link layer is very well tracked. Sharing such information with the network layer can significantly improve mobility management. Thus, cross-layer mechanisms need to be devised to also consider edge servers mobility. Our service migration approach is a beginning in this direction.

\subsection {Dynamic Resource Management} Enabling a PEC framework requires fast setup and tear-down of routes to resources joining and leaving the PEC pool, respectively. 
Scalable cross-layer protocols need to be designed to perform this efficiently. 
Moreover, there is a possibility that accessing the back-end cloud is not possible given the critical needs of the application(s). For instance, in disaster scenarios, a surviving smart city infrastructure can/has to operate without connection to the central cloud. The distributed control plane can help the PEC continue operations. Particularly, several PECs can join together to help in essential post-disaster communications. The PEC design should be upgraded to enable seamless transition.
    
\subsection {Security, Privacy, and Availability} PEC's highly dynamic and multi-stakeholder nature open doors for new attacks from all the stakeholders. Malicious entities can deploy fake access infrastructure (\eg base stations or access points) to hijack communication for blackhole or wormhole attacks. Fake or compromised orchestrators may advertise compromised services to obtain private information or interrupt service delivery. A malicious service from the application providers can alter the result of a service execution, leak private information, provide a backdoor to malicious users, escalate privileges, or compromise the computing entity. Similarly, malicious data from the users may include Trojans, viruses, or malware aimed at interrupting service for others. In addition to these threats, PEC infrastructure will be vulnerable to classical attacks, such as DDoS attacks on the edge infrastructure to exhaust the edge infrastructure resources and service overcharging by greedy orchestrators.
    
The majority of the existing security measures are designed for more static and single domain environments, resulting in their ineffectiveness in addressing PEC's threats. For instance the existing off-premise and cloud-based authentication and authorization frameworks, including OAuth framework, are unable to perform access control delegation to the third-party EC infrastructure. Thus, security, identity management, and trust mechanisms need to be re-visited taking PEC's dynamicity and multi-stakeholders nature into account. 

%
\section{Concluding Remarks}
\label{Conclusion}
We presented the concept of Pervasive Edge Computing (PEC) and design an NDN-based architecture for PEC, which we foresee as the future of EC. We discussed intra- and inter-stakeholder service orchestration, including service deployment, discovery, invocation, and migration. We also discussed the PEC economic model. We elaborated on the challenges that need to be addressed for successful deployment of our envisioned PEC framework and further discussed potential opportunities and areas of research.     
%
\bibliographystyle{IEEEtran}
\bibliography{paper}
%
%
\begin{IEEEbiography}[{\includegraphics[width=1in,clip,keepaspectratio]{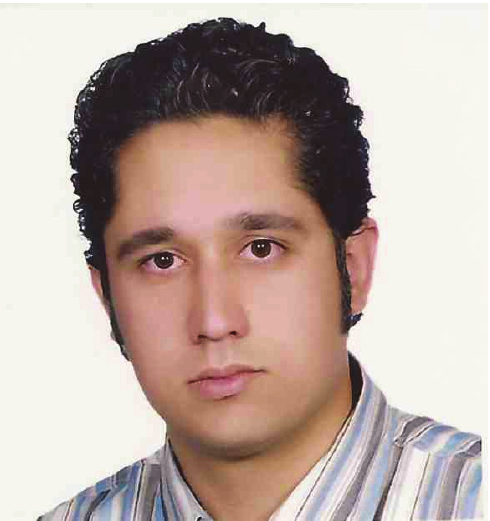}}]{Reza Tourani}
(GSM'12, M'18) is an assistant professor in computer science at Saint Louis University. He received Ph.D. and M.S. in computer science from NMSU, NM, USA, in 2018 and 2012, respectively.  His research interests are in the areas of security and privacy, FIA, Cyber-physical systems, smart grid, IoT, and edge computing. He has authored more than 25 peer-reviewed IEEE/ACM journal articles and conference proceedings. 
\end{IEEEbiography}
\begin{IEEEbiography}[{\includegraphics[width=1in,clip,keepaspectratio]{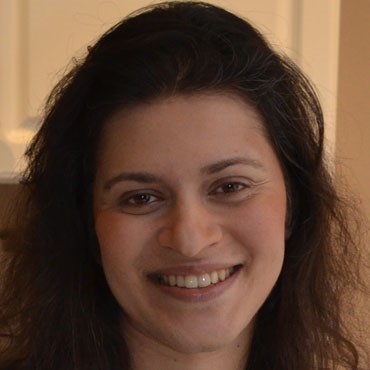}}]{Srikathyayani Srikanteswara (Kathyayani)} is a Sr. Research Scientist at Intel Labs leading research in ICN for Edge Computing and spectrum sharing. She helped develop Intel's policy on shared spectrum and lead standardization efforts in the US and Europe. Prior to Intel, she was with Navsys Corporation, working on advanced GPS receiver techniques using SDR, and was a research faculty member at Virginia Tech. She received her MS and Ph.D. in Electrical Engineering from Virginia Tech.
\end{IEEEbiography}
\begin{IEEEbiography}[{\includegraphics[width=1in,clip,keepaspectratio]{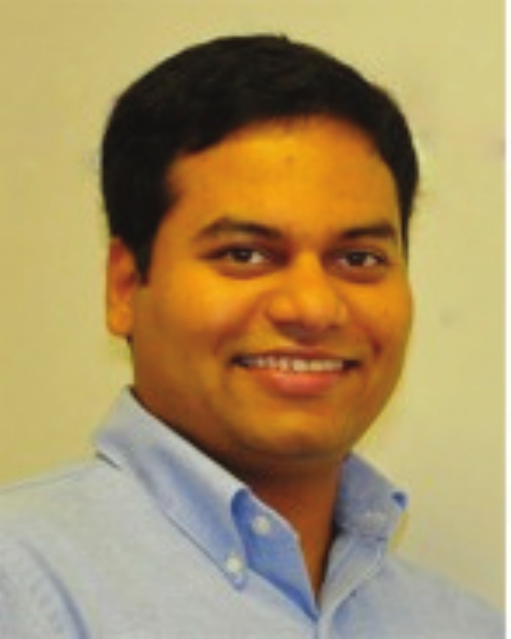}}]{Satyajayant Misra}
(SM'05, M'09) is an associate professor in computer science at New Mexico State University. He completed his Ph.D. in Computer Science from ASU, AZ, USA, in 2009. His research interests are in wireless networks, the Internet, and smart grid architectures and protocols. He served on several IEEE/ACM journal editorial boards and conference executive committees including editorship in IEEE IoT Journal and IEEE Wireless Communication Magazine. He has authored over 80 peer-reviewed publications. 
\end{IEEEbiography}
\begin{IEEEbiography}[{\includegraphics[width=1in,clip,keepaspectratio]{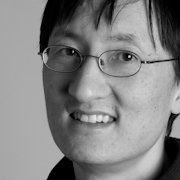}}]{Richard Chow}
is a University Research Director and Scientist at Intel Corporation. Prior to Intel, he has held positions at PARC, Samsung Electronics R\&D, Yahoo, and Motorola. His work concentrates on privacy, big data, and cloud. He has over 20 patents and 30 peer-reviewed journals, conference papers, and book chapters. He was awarded runner-up for the 2010 PET Award for Outstanding Research in Privacy Enhancing Technologies. He has given invited talks RSA, BlackHat, and OWASP. He has a Ph.D. in mathematics from UCLA.
\end{IEEEbiography}
\begin{IEEEbiography}[{\includegraphics[width=1in,clip,keepaspectratio]{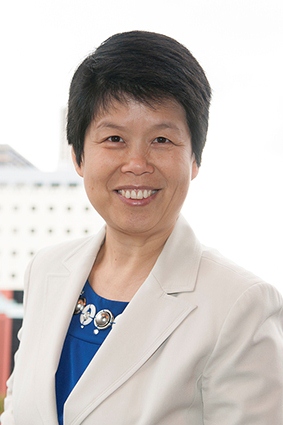}}]{Lily Yang} is a Principal Engineer in Security and Privacy Research at Intel Labs, USA. Lily holds 17 patents in data compression, video and image coding, wireless communications, and security for Cyber-Physical systems. She has published 20+ research and technical papers and contributed to industry and international standards such as IETF, IEEE, WiGig, ETSI. Lily holds a Ph.D. in Electrical Engineering from Washington State University.
\end{IEEEbiography}
\begin{IEEEbiography}[{\includegraphics[width=1in,clip,keepaspectratio]{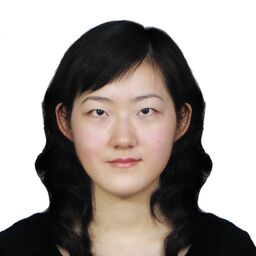}}]{Xiruo Liu} is currently a Research Scientist with Security and Privacy Research Lab at Intel Labs. Xiruo holds a Ph.D. degree in Electrical and Computer Engineering from Rutgers University. Her research interests include network security and privacy with focuses on vehicular communications, 5G, IoT, and FIA. She also participates in related industry standardization activities.
\end{IEEEbiography}
\begin{IEEEbiography}[{\includegraphics[width=1in,clip,keepaspectratio]{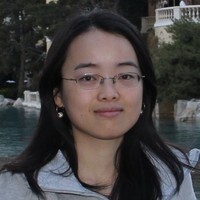}}]{Yi Zhang} is a Research Scientist in Wireless Communication Research at Intel Labs. In the past, she held positions as Senior Researcher at Nokia and Senior System Engineer at NetScout. Yi has 50+ patents on Wireless Communication and ICN. Yi holds a Ph.D. in Electrical Engineering from Beijing University of Posts and Communications. 
\end{IEEEbiography}
\end{document}